\documentclass[aps,prl,twocolumn,superscriptaddress,showpacs,nofootinbib,floatfix,10pt]{revtex4-1}
\usepackage{graphicx,dcolumn,amsmath,verbatim,rotating,bm,textcomp,color}
\newcolumntype{x}{D{/}{/}{2.2}}
\newcolumntype{y}{D{.}{.}{4.5}}
\newcolumntype{z}{D{)}{)}{6.0}}
\newcolumntype{w}{D{)}{)}{7.4}}
\definecolor{darkblue}{RGB}{46,48,146}
\usepackage{ifpdf}
\ifpdf
\usepackage[
	pdfpagemode=UseNone,
  colorlinks=true,
	linkcolor=darkblue,
	filecolor=darkblue,
	urlcolor=darkblue,
	citecolor=darkblue,
	bookmarks=false,
	pdftex,
	plainpages=false,
	hypertexnames=false,
	pdfpagelabels=true,
	hyperindex=true]{hyperref}
\DeclareGraphicsExtensions{.pdf}
\else
\DeclareGraphicsExtensions{.eps}
\fi

\begin{document}

\title{Spins, Electromagnetic Moments, and Isomers of $^\textbf{107-129}$Cd}

\author{\mbox{D.~T.~Yordanov}}
\email[]{\mbox{Deyan.Yordanov@cern.ch}}
\affiliation{\mbox{Max-Planck-Institut f\"ur Kernphysik, Saupfercheckweg 1, D-69117 Heidelberg, Germany}}
\affiliation{\mbox{CERN European Organization for Nuclear Research, Physics Department, CH-1211 Geneva 23, Switzerland}}

\author{\mbox{D.~L.~Balabanski}}
\affiliation{INRNE, Bulgarian Academy of Science, BG-1784 Sofia, Bulgaria}

\author{\mbox{J.~Biero\'n}}
\affiliation{\mbox{Instytut Fizyki imienia~Mariana~Smoluchowskiego, Uniwersytet Jagiello\'nski, Reymonta 4, 30-059~Krak\'ow, Poland}}

\author{\mbox{M.~L.~Bissell}}
\affiliation{\mbox{Instituut voor Kern- en Stralingsfysica, KU Leuven, Celestijnenlaan 200D, B-3001 Leuven, Belgium}}

\author{\mbox{K.~Blaum}}
\affiliation{\mbox{Max-Planck-Institut f\"ur Kernphysik, Saupfercheckweg 1, D-69117 Heidelberg, Germany}}

\author{\mbox{I.~Budin\v{c}evi\'c}}
\affiliation{\mbox{Instituut voor Kern- en Stralingsfysica, KU Leuven, Celestijnenlaan 200D, B-3001 Leuven, Belgium}}

\author{\mbox{S.~Fritzsche}}
\affiliation{\mbox{GSI Helmholtzzentrum f\"ur Schwerionenforschung GmbH, D-64291 Darmstadt, Germany}}

\author{\mbox{N.~Fr\"ommgen}}
\affiliation{\mbox{Institut f\"ur Kernchemie, Johannes Gutenberg-Universit\"at Mainz, D-55128 Mainz, Germany}}

\author{\mbox{G.~Georgiev}}
\affiliation{CSNSM-IN2P3-CNRS, Universit\'e de Paris Sud, F-91405 Orsay, France}

\author{\mbox{Ch.~Geppert}}
\affiliation{\mbox{GSI Helmholtzzentrum f\"ur Schwerionenforschung GmbH, D-64291 Darmstadt, Germany}}
\affiliation{\mbox{Institut f\"ur Kernchemie, Johannes Gutenberg-Universit\"at Mainz, D-55128 Mainz, Germany}}

\author{\mbox{M.~Hammen}}
\affiliation{\mbox{Institut f\"ur Kernchemie, Johannes Gutenberg-Universit\"at Mainz, D-55128 Mainz, Germany}}

\author{\mbox{M.~Kowalska}}
\affiliation{\mbox{CERN European Organization for Nuclear Research, Physics Department, CH-1211 Geneva 23, Switzerland}}

\author{\mbox{K.~Kreim}}
\affiliation{\mbox{Max-Planck-Institut f\"ur Kernphysik, Saupfercheckweg 1, D-69117 Heidelberg, Germany}}

\author{\mbox{A.~Krieger}}
\affiliation{\mbox{Institut f\"ur Kernchemie, Johannes Gutenberg-Universit\"at Mainz, D-55128 Mainz, Germany}}

\author{\mbox{R.~Neugart}}
\affiliation{\mbox{Institut f\"ur Kernchemie, Johannes Gutenberg-Universit\"at Mainz, D-55128 Mainz, Germany}}

\author{\mbox{W.~N\"ortersh\"auser}}
\affiliation{\mbox{GSI Helmholtzzentrum f\"ur Schwerionenforschung GmbH, D-64291 Darmstadt, Germany}}
\affiliation{\mbox{Institut f\"ur Kernchemie, Johannes Gutenberg-Universit\"at Mainz, D-55128 Mainz, Germany}}

\author{\mbox{J.~Papuga}}
\affiliation{\mbox{Instituut voor Kern- en Stralingsfysica, KU Leuven, Celestijnenlaan 200D, B-3001 Leuven, Belgium}}

\author{\mbox{S.~Schmidt}}
\affiliation{\mbox{GSI Helmholtzzentrum f\"ur Schwerionenforschung GmbH, D-64291 Darmstadt, Germany}}

\date{\today}
\begin{abstract}

The neutron-rich isotopes of cadmium up to the $N=82$ shell closure have been investigated by high-resolution laser spectroscopy. Deep-UV excitation at $214.5$~nm and radioactive-beam bunching provided the required experimental sensitivity. Long-lived isomers are observed in $^{127}$Cd and $^{129}$Cd for the first time. One essential feature of the spherical shell model is unambiguously confirmed by a linear increase of the $11/2^-$ quadrupole moments. Remarkably, this mechanism is found to act well beyond the $h_{11/2}$ shell.

\end{abstract}
\maketitle

When first proposed the nuclear shell model was largely justified on the basis of magnetic-dipole properties of nuclei \cite{Mayer}. The electric quadrupole moment could have provided an even more stringent test of the model, as it has a very characteristic linear behavior with respect to the number of valence nucleons \cite{MayerJensen,HorieArimaQ}. However, the scarcity of experimental quadrupole moments at the time did not permit such studies. Nowadays, regardless of experimental challenges, the main difficulty is to predict which nuclei are likely to display this linear signature. The isotopes of cadmium, investigated here, proved to be the most revealing case so far. Furthermore, being in the neighborhood of the ``magic'' tin, cadmium is of general interest for at least two additional reasons. First, theory relies on nuclei near closed shells for predicting other, more complex systems. Second, our understanding of stellar nucleosynthesis strongly depends on the current knowledge of nuclear properties in the vicinity of the doubly magic tin isotopes \cite{Hakala}. Moreover, specific questions concerning the nuclear structure of the cadmium isotopes require critical evaluation, such as: shell quenching \cite{Kautzsch,Dillmann}, sphericity \cite{Dworschak}, deformation \cite{Jungclaus,Rodriguez}; or whether vibrational nuclei exist at all \cite{GarrettWood}. Some of these points will be addressed here quite transparently, others require dedicated theoretical work to corroborate our conclusions. In this Letter we report advanced measurements by collinear laser spectroscopy on the very neutron-rich cadmium isotopes. Electromagnetic moments in these complex nuclei are found to behave in an extremely predictable manner. Yet, their description goes beyond conventional interpretation of the nuclear shell model.

The measurements were carried out with the collinear laser spectroscopy setup at ISOLDE-CERN. High-energy protons impinging on a tungsten rod produced low- to medium-energy neutrons inducing fission in a uranium carbide target. Proton-rich spallation products, such as cesium, were largely suppressed in this manner. Further reduction of surface-ionized isobaric contamination was achieved by the use of a quartz transfer line \cite{UKoster}, which allowed the more volatile cadmium to diffuse out of the target while impurities were retained sufficiently long to decay. Cadmium atoms were laser ionized, accelerated to an energy of $30$~keV, and mass separated. The ion beam was injected into a gas-filled radio-frequency Paul trap \cite{EMane} and extracted typically every 100~ms as short bunches with a temporal width of about 5~$\mu\text{s}$. The ratio of the above time constants equals the factor of background suppression and therefore results in an increase of the overall sensitivity by the square root of that factor ($\approx10^2$).

The ion of cadmium was excited in the transition: $5s\;^2S_{1/2}\rightarrow 5p\;^2P_{3/2}$ at $214.5$~nm. Continuous-wave laser light of this wavelength was produced by sequential second-harmonic generation from the output of a titanium-sapphire laser, pumped at $532$~nm. The combined fourth-harmonic generation is characterized by a conversion efficiency of up to $2$~\%.
Optimal laser power of about 1~mW was supplied for the measurements. Using the ion for laser excitation increased the overall sensitivity by more than an order of magnitude with respect to the neutral atom. The improvement can be accounted for by the faster transition, the higher quantum efficiency of detection, and the absence of ion-beam neutralization. Such establishment of deep-UV laser beams could potentially provide access to isotopic chains thus far unstudied due to demanding transition wavelengths.

In the conventional manner the atomic hyperfine structure was detected by the ion-beam fluorescence as a function of the laser frequency scanned via the Doppler effect. This method is to a large extent insensitive to contaminant beams. However, care has been taken not to exceed $10^6$ ions accumulated in the Paul trap in order to avoid space-charge effects. This condition was not a limiting factor for the experiment.

\begin{figure}[t]
\begin{center}
\includegraphics[angle=0,width=\linewidth]{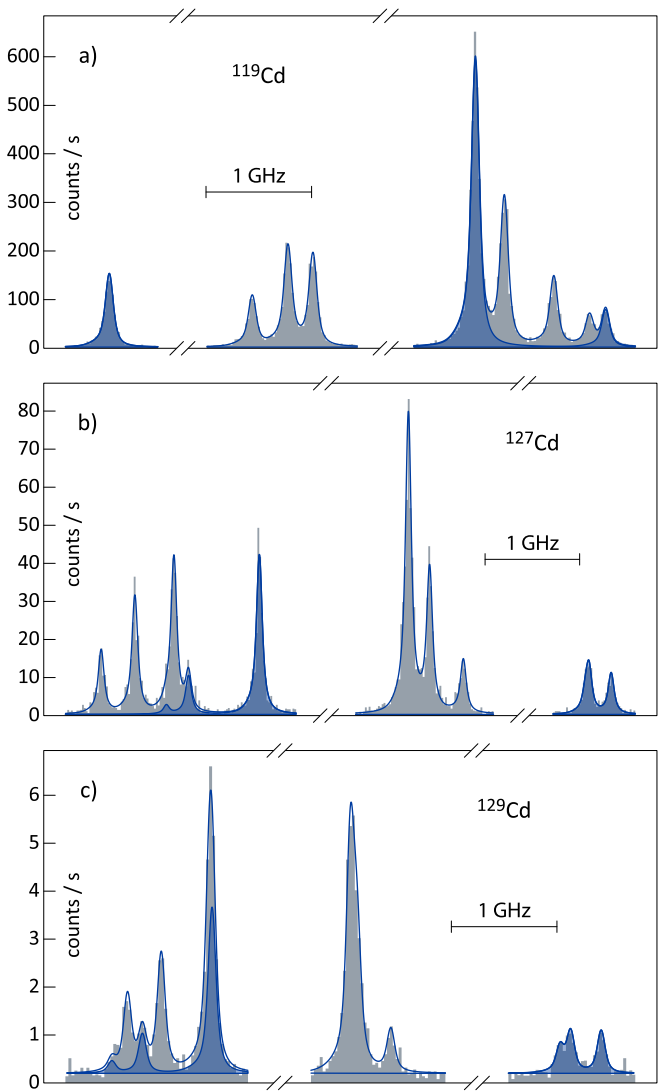}
\caption{\label{fig1}
Example hfs spectra of $^{119}$Cd (a), $^{127}$Cd (b), and $^{129}$Cd (c). Only frequency regions containing hfs components are displayed. The fitted curve incorporates two states on a common background. The lower-spin state is indicated by a semi-transparent fill.}
\end{center}
\end{figure}
An important accomplishment of this work is the discovery of long-lived isomers in $^{127}$Cd and $^{129}$Cd. Representative spectra are displayed in Figs.~\ref{fig1}~(b) and (c) where the presence of two nuclear states is clearly identified. It is impossible to determine from the optical measurements alone which of the two is the ground state and what their respective decay modes are. Spins and electromagnetic moments, on the other hand, were determined successfully for both states in each of the isotopes. The presence of such isomers has been suggested in previous studies \cite{Hoteling,Naqvi,Kratz}.

The experimental results are presented in Tab.~\ref{tab1}. Some comments on the spin measurements apply here. The hyperfine structure clearly identifies a ground-state spin of $5/2$ for $^{107}$Cd and $^{109}$Cd. Spin $1/2$ is assigned to all ground states from $^{111}$Cd to $^{119}$Cd due to the reduced number of hfs components, three instead of six. A typical example is $^{119}$Cd in Fig.~\ref{fig1}~(a), whose spin adopted in the literature \cite{ndsa119} is therefore incorrect. The $3/2$ assignments in $^\text{121-129}$Cd are strongly supported by $\chi^2$ analysis of relative hfs intervals and line intensities. Furthermore, the magnetic moments are consistent with an odd-neutron occupation of the $d_{3/2}$ orbital. The hyperfine structure offers limited sensitivity to high spins. Nevertheless, all $11/2$ assignments are rather firm, since the corresponding electromagnetic moments in Fig.~\ref{fig2} are clearly of $h_{11/2}$ origin.

The $S_{1/2}$ hyperfine parameters $A$ are measured with precision on the level of detectable hyperfine anomaly. Accurate results were deduced with the following procedures. For the observed spins of $1/2$, $5/2$ and $11/2$, there are isotopes in the cadmium chain studied by NMR. The hyperfine anomaly within a set of states with identical spins was neglected and each set was assigned a high-precision value of the corresponding spin as a reference. The resulting magnetic moments are in good agreement with NMR measurements, as evident from Tab.~\ref{tab1}. For the $3/2$ magnetic moments a hyperfine-anomaly correction was applied with the semi-empirical approach of Moskowitz and Lombardi \cite{Moskowitz}:
\begin{equation}
\label{eq1}
\frac{A}{A_0}\cdot\frac{I}{I_0}\cdot\frac{\mu_0}{\mu}-1=\frac{\alpha}{|\mu_0|}-\frac{\alpha}{|\mu|}\cdot
\end{equation}

Quadrupole moments were derived from the hyperfine parameters $B$ using the relation: $B=e\,Q\,V_{JJ}$, where $V_{JJ}$ is the electric field gradient at the nucleus and $e$ is the electron charge. Dirac-Hartree-Fock \cite{GrantBook} calculations provided the field gradient in the $5p\;^2P_{3/2}$ state of the cadmium ion. The finite-difference code GRASP \cite{grasp2K} was used to generate the numerical-grid wave functions in conjunction with tools and methodology for hyperfine-structure applications previously described \cite{BieronBeF,BieronHg,BieronAu}. The theoretical error bar was evaluated by applying several methods of orbital generation. Details on the applied computational procedure will be published elsewhere. The obtained electric field gradient is presented in Tab.~\ref{tab1} along with the quadrupole moments thus determined independent of previous studies. Note that the literature values are about $14$~\%
larger in magnitude as they are all referenced to a semi-empirical calculation of the electric field gradient for $^{109}$Cd \cite{Laulainen}. Much of this discrepancy can be accounted for by the Sternheimer shielding, which is intrinsically included in our calculation, but not in the above-mentioned work.

The linear behavior of the $11/2^-$ quadrupole moments is the most striking and revealing feature of the cadmium nuclei. Moreover, the trend is found to persist uninterruptedly over a sequence of $10$ odd-mass isotopes, as shown in Fig.~\ref{fig2}~(b). The most likely cause behind this phenomenon is the unique parity of the $h_{11/2}$ orbital, which would favor simpler shell-model configurations for the $11/2^-$ states. Indeed, in the $jj$-coupling shell model by Mayer and Jensen \cite{MayerJensen} single-shell proton states would exhibit a linear increase with respect to the number of protons. Horie and Arima justified a similar dependence for neutrons \cite{HorieArimaQ} by taking into account their interaction with protons. In a more general sense, we consider here the seniority scheme and its formalism by de-Shalit and Talmi \cite{ShalitTalmi}:
\begin{figure}[t]
\begin{center}
\includegraphics[angle=0,width=\linewidth]{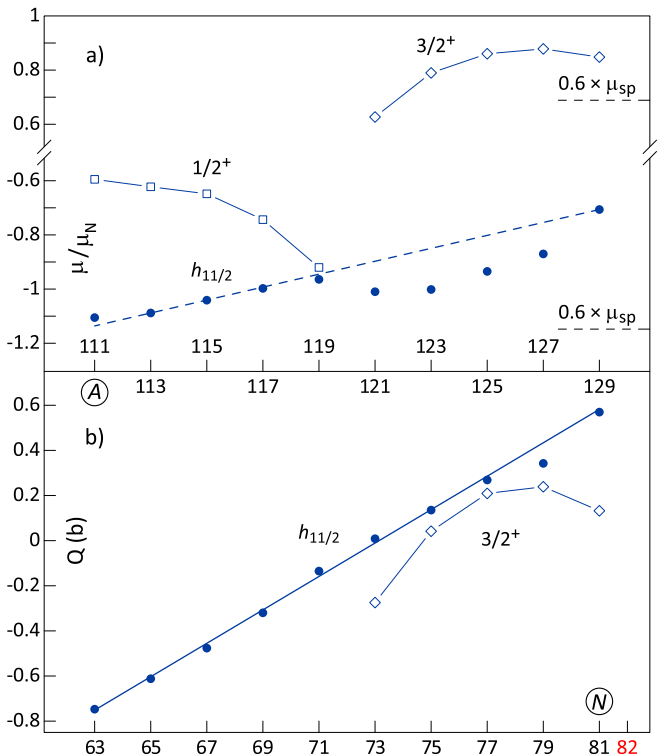}
\caption{\label{fig2}
Magnetic (a) and quadrupole (b) moments of $^{111\text{-}129}$Cd from this work. The experimental error bars are smaller than the markers. A straight line is fitted through the $h_{11/2}$ quadrupole moments, consistent with Eq.~(\ref{eq2}). The dashed line indicates the effect of core polarization.}
\end{center}
\end{figure}
\begin{table*}[t]
\caption{\label{tab1}
Spins, hyperfine parameters, and electromagnetic moments derived from this work. Experimental uncertainties (uncorrelated) are quoted in parentheses. Uncertainties on the quadrupole moments due to the electric field gradient (correlated) are enclosed in square brackets. Correction for the hyperfine anomaly is applied to the magnetic moments by using separate NMR references for the states with spin $1/2$, $5/2$, and $11/2$, and by the Moskowitz-Lombardi rule \cite{Moskowitz} for the states with spin $3/2$. High-precision magnetic moments calculated from NMR frequency ratios \cite{Chaney,Spence} relative to the proton \cite{codata} and corrected for diamagnetism \cite{Raghavan} are displayed for comparison.}
\begin{ruledtabular}
\begin{tabular}{cxydrlzw}
\multicolumn{1}{c}{$Z+N$}
    & \multicolumn{1}{c}{$I$}
           & \multicolumn{1}{c}{$A_{(5p\;^2P_{3/2})}$~(MHz)}
                            & \multicolumn{1}{c}{$A_{(5s\;^2S_{1/2})}$~(MHz)}
                                               & \multicolumn{1}{c}{$\mu$~$(\mu_\text{N})$}
                                                                & \multicolumn{1}{c}{$\mu_\text{(NMR)}$~$(\mu_\text{N})$}
                                                                                                                                  & \multicolumn{1}{c}{$B_{(5p\;^2P_{3/2})}$~(MHz)}
                                                                                                                                              & \multicolumn{1}{c}{$Q$~(mb)} \\
\hline
107 &  5/2 &  -82.3\,~\,(3) &  -3009.8\,~\,(7) & $-0.6151\,(2)$ & $-0.6150554\,  (11)$                                            &  401\,(2) &  601\,~\,(3)  [24] \\
109 &  5/2 & -111.4\,~\,(2) &  -4051.0\,~\,(7) &                & $-0.8278461\,  (15)$\footnotemark[1]                            &  403\,(1) &  604\,~\,(1)  [25] \\
111 &  1/2 & -398.2\,~\,(5) & -14535.0\,  (23) &                & $-0.5948861\,~\,(8)$\footnotemark[1]$^\text{,}$\footnotemark[2] &           &                    \\
111 & 11/2 &  -67.2\,~\,(2) &  -2456.9\,~\,(5) & $-1.1052\,(3)$ &                                                                 & -498\,(3) & -747\,~\,(4)  [30] \\
113 &  1/2 & -418.5\,~\,(6) & -15208.0\,  (23) & $-0.6224\,(2)$ & $-0.6223009\,~\,(9)$                                            &           &                    \\
113 & 11/2 &  -66.4\,~\,(2) &  -2419.3\,~\,(6) & $-1.0883\,(3)$ & $-1.0877842\,  (17)$                                            & -408\,(2) & -612\,~\,(3)  [25] \\
115 &  1/2 & -434.1\,  (10) & -15840.6\,  (30) & $-0.6483\,(2)$ & $-0.6484259\,  (12)$                                            &           &                    \\
115 & 11/2 &  -63.7\,~\,(2) &  -2314.2\,~\,(4) &                & $-1.0410343\,  (15)$\footnotemark[1]                            & -317\,(3) & -476\,~\,(5)  [19] \\
117 &  1/2 & -499.2\,  (11) & -18168.5\,  (32) & $-0.7436\,(2)$ &                                                                 &           &                    \\
117 & 11/2 &  -60.8\,~\,(3) &  -2217.5\,~\,(8) & $-0.9975\,(4)$ &                                                                 & -213\,(4) & -320\,~\,(6)  [13] \\
119 &  1/2 & -615.5\,  (13) & -22482.0\,  (39) & $-0.9201\,(2)$ &                                                                 &           &                    \\
119 & 11/2 &  -59.0\,~\,(2) &  -2143.3\,~\,(4) & $-0.9642\,(3)$ &                                                                 &  -90\,(2) & -135\,~\,(3)~\,[5] \\
121 &  3/2 &  139.7\,  (15) &   5106.2\,  (34) & $ 0.6269\,(7)$ &                                                                 & -183\,(5) & -274\,~\,(7)  [11] \\
121 & 11/2 &  -62.0\,~\,(3) &  -2245.3\,~\,(8) & $-1.0100\,(4)$ &                                                                 &    6\,(4) &    9\,~\,(6)       \\
123 &  3/2 &  175.5\,  (13) &   6435.6\,  (27) & $ 0.7896\,(6)$ &                                                                 &   28\,(3) &   42\,~\,(5)~\,[2] \\
123 & 11/2 &  -61.7\,~\,(2) &  -2226.3\,~\,(5) & $-1.0015\,(3)$ &                                                                 &   90\,(3) &  135\,~\,(4)~\,[6] \\
125 &  3/2 &  193.5\,~\,(7) &   7012.6\,  (19) & $ 0.8603\,(6)$ &                                                                 &  139\,(3) &  209\,~\,(4)~\,[9] \\
125 & 11/2 &  -57.0\,~\,(2) &  -2077.9\,~\,(4) & $-0.9347\,(2)$ &                                                                 &  179\,(5) &  269\,~\,(7)  [11] \\
127 &  3/2 &  195.3\,  (12) &   7159.6\,  (31) & $ 0.8783\,(7)$ &                                                                 &  159\,(3) &  239\,~\,(5)  [10] \\
127 & 11/2 &  -52.6\,~\,(3) &  -1934.5\,~\,(5) & $-0.8702\,(3)$ &                                                                 &  228\,(7) &  342\,  (10)  [14] \\
129 &  3/2 &  187.7\,  (23) &   6912.9\,  (48) & $ 0.8481\,(8)$ &                                                                 &   88\,(5) &  132\,~\,(7)~\,[5] \\
129 & 11/2 &  -44.1\,~\,(5) &  -1570.2\,  (11) & $-0.7063\,(5)$ &                                                                 &  380\,(9) &  570\,  (13)  [23] \\
\hline
\multicolumn{8}{r}{Electric field gradient: $e\,V_{JJ}/h=666\,(27)$~(MHz/b)} \\
\end{tabular}
\end{ruledtabular}
\footnotetext[1]{Magnetic moment used as a reference for the states with the corresponding spin ($\mu_0$ in Eq.~(\ref{eq1}), with $\alpha=0$~$\mu_\text{N}$).}
\footnotetext[2]{$\mu_0$ for the $3/2$ states ($\alpha/\mu_\text{N}=1.7$~{\textperthousand}). The experimental uncertainties of the $3/2$ magnetic moments are convoluted with $5\times10^{-4}$~$\mu_\text{N}$ representing the standard deviation of the scatter when the hyperfine anomaly is neglected and different reference values are used.}
\end{table*}
\begin{equation}
\label{eq2}
\langle j^n|\hat{Q}| j^n \rangle=\frac{2j+1-2n}{2j+1-2\nu}\langle j^\nu|\hat{Q}| j^\nu \rangle\cdot
\end{equation}
The origin of Eq.~(\ref{eq2}) is easier to understand in the particular case of seniority $\nu=1$, or ``normal coupling'' \cite{Mayer,MayerJensen}, when all but one particle are coupled to spin zero. By definition the quadrupole moment corresponds to the state with maximum angular-momentum projection, therefore the magnetic substates $m=\pm j$ are not available for nucleon pairs. This will produce a quadrupole moment dependent on the number of nucleons $n$. Since the number of $j^n$ configurations is $(2j+1)/2$, Eq.~(\ref{eq2}) could explain the alignment of only $6$ quadrupole moments for spin $11/2$. Furthermore, the possibility of configurations with different seniorities following the same trend can be excluded. For instance, the matrix element $\langle j^\nu|\hat{Q}| j^\nu \rangle$ for seniority $\nu=3$, calculated with the aid of tabulated coefficients of fractional parentage \cite{BaymanLande}, is $-8$~\%
of the single-particle quadrupole moment $Q_\text{sp}=\langle j|\hat{Q}| j \rangle$. Such values would greatly deviate from the experimental trend. Clearly, one has to surrender the integer nature of $n$ and interpret it as the actual neutron occupation. This is possible if one assumes that the population of neutron pairs ($I=0$) is shared between the neighboring orbitals: $s_{1/2}$, $d_{3/2}$, $d_{5/2}$, and $h_{11/2}$, suggesting a kind of degeneracy in terms of total energy per pair. The odd particle, on the other hand, must always occupy $h_{11/2}$, as migration to any other orbital in the shell would result in a change of parity. Finally, assuming no particle-hole excitations across $N=82$, one can substitute: $n=1+p(A-n_0)$, where $n_0=111$, $A=N+Z$, and $p=5/9$. The probability $p$ for pair occupation of $h_{11/2}$ is calculated as the capacity of $h_{11/2}$ for neutron pairs in addition to an odd neutron, divided by the number of pairs filled between $^{111}$Cd and $^{129}$Cd. It can be easily verified that with this substitution there is exactly one $h_{11/2}$ neutron in $^{111}$Cd and eleven in $^{129}$Cd. An examination of Eq.~(\ref{eq2}) shows that the quadrupole moments should cross zero in the middle of the shell, which in the current description corresponds to $A=120$. Indeed, the crossing point was determined at $^{121}$Cd, very close to that prediction. In order to account for the small deviation of one mass unit, the data in Fig.~\ref{fig2}~(b) are fitted with an offset term $Q_\text{const}$ representing a constant quadrupole-moment contribution from correlations with the core. The resulting fit parameters are: $Q_\text{sp}=-667(31)$~mb and $Q_\text{const}=-85(8)$~mb. For comparison, the single-particle quadrupole moment of $h_{11/2}$ can be estimated by: $-\langle r^2\rangle(2j-1)/(2j+2)=-269$~mb. Here, under the assumption of a uniformly charged spherical nucleus, the mean square radius of the last orbital is approximated by $5/3$ of the mean square charge radius of $^{111}$Cd \cite{FrickeHeilig}. The ratio of the two values implies a relatively large effective charge $e_n=2.5e$. This result is commented on below in connection with the magnetic moments. The line of quadrupole moments crossing zero essentially in the middle of the $h_{11/2}$ shell indicates a spherical shape for the $11/2^-$ states. However, one has to acknowledge the deviation from the straight line at $^{127}$Cd. It is a small negative effect occurring between $^{126}$Cd and $^{128}$Cd, for which abnormal first $2^+$ energies are reported \cite{Kautzsch}. The meaning of this observation should be further evaluated in light of possible shell quenching \cite{Kautzsch,Dillmann} against suggested deformation \cite{Jungclaus,Rodriguez}. The $n$ dependence of the nuclear quadrupole moment has been investigated in the past \cite{FWByron} and more recently reviewed in the $i_{13/2}$ isomers of lead and mercury \cite{Neyens}. The results reported here are the first to demonstrate persistence of that mechanism beyond a single shell. Furthermore, the exceptionally high experimental precision achieved here provides a far more stringent definition of a linear trend.

The nuclei of cadmium exhibit yet another simple trend. Their $11/2^-$ magnetic moments, as shown in Fig.~\ref{fig2}~(a), increase linearly from $^{111}$Cd to $^{129}$Cd. Four isotopes in the range $^{121\text{-}127}$Cd make an exception, which appears to be correlated with the spin change of the second long-lived state. Seemingly, this linear dependence is inconsistent with our description of the quadrupole moments since any odd number of nucleons in a single shell would produce the same magnetic moment as a single nucleon \cite{Mayer,MayerJensen}. In this respect one may consider $^{129}$Cd where all neutron orbitals apart from a single $h_{11/2}$ hole are fully occupied with no apparent possibility of ``configuration mixing'' \cite{ArimaHorieM}. It is then expected that the $11/2^-$ magnetic moment of $^{129}$Cd should be the most consistent one with the single-particle value, yet it deviates the most. Clearly, an accurate description of the cadmium isotopes should account for the two holes in the $Z=50$ proton core. First-order core polarization does indeed generate a linear $n$ dependence of the magnetic moment \cite{Nomura}, though higher-order contributions may be important as well \cite{Wouters}. The quadrupole moments, on the other hand, are influenced by this proton-core polarization only through the effective charge, whose large value can now be understood.

In summary, advanced laser spectroscopy provided access to the very exotic odd-mass isotopes of cadmium within the $N=82$ shell. Long-lived $11/2^-$ states are identified in $^{127}$Cd and $^{129}$Cd for the first time. Remarkably, all quadrupole moments associated with the unique-parity $h_{11/2}$ orbital increase linearly with respect to the number of neutrons, as predicted by the extreme shell-model. Yet, this linear trend is found to extend well beyond the single $h_{11/2}$ shell. Interpretation of both magnetic and quadrupole moments is offered in simple terms and in a common framework.

\begin{acknowledgments}

This work has been supported by the Max-Planck Society, the German Federal Ministry for Education and Research under contract no. 05P12RDCIC, the Belgian IAP project P7/12, the FWO-Vlaanderen, and the European Union seventh framework through ENSAR under contract no. 262010. The atomic calculations were supported by the Polish National Center of Science (N~N202~014140). N.~F. is a fellowship recipient through GRK Symmetry Breaking (DFG/GRK~1581). D.~L.~B. acknowledges support from the Bulgarian National Science Fund (DID-02/16). We thank the ISOLDE technical group for their professional assistance.

\end{acknowledgments}

\end{document}